\documentclass{article}

\usepackage{arxiv}
\usepackage[utf8]{inputenc} 
\usepackage[T1]{fontenc}    
\usepackage{hyperref}       
\usepackage{booktabs}       
\usepackage{amsfonts}       
\usepackage{amsmath}       
\usepackage{nicefrac}      
\usepackage{microtype}      
\usepackage{graphicx}
\usepackage{natbib}
\usepackage{doi}
\usepackage{xcolor}
\usepackage{amsmath}
\usepackage{textcomp}

\title{Principles of Use of Tensile J-Curve Materials in Antagonistic Arrangements}

\usepackage{authblk}
\author[a]{Liuyang Cheng}
\author[c]{Wonsik Eom}
\author[a]{Qiong Wang}
\author[a]{Hyeongkeun Kim}
\author[d]{Roberto Pineda Guzman}
\author[a]{Jeongmin Kim}
\author[b]{Montse Solis}
\author[a]{Shreyas Malladi}
\author[a]{Samuel Tsai}
\author[a,e,f]{Mariana E. Kersh}
\author[a,f,1]{Sameh H. Tawfick}

\affil[a]{Department of Mechanical Science and Engineering, The Grainger College of Engineering, University of Illinois Urbana-Champaign, Urbana, IL 61801}
\affil[b]{Department of Materials Science and Engineering, The Grainger College of Engineering, University of Illinois Urbana-Champaign, Urbana, IL 61801}
\affil[c]{Department of Fiber Convergence Materials Engineering, Dankook University, Yongin-si, Republic of Korea 16890}
\affil[d]{Carle Clinical Imaging Research Program, Stephens Family Clinical Research Institute, Carle Health, Urbana, IL 61801}
\affil[e]{Carle Illinois College of Medicine, University of Illinois Urbana-Champaign, Urbana, IL 61801}
\affil[f]{Beckman Institute for Advanced Science and Technology, University of Illinois Urbana-Champaign, Urbana, IL 61801}

\usepackage{xwatermark}
\newwatermark[firstpage,color=gray!90,angle=0,scale=0.28, xpos=0in,ypos=-5in]{\textsuperscript{1}To whom correspondence should be addressed. E-mail: \texttt{tawfick@illinois.edu} }

\begin{document}
\maketitle

\begin{abstract}
	Natural ligaments are soft connective tissues that must simultaneously provide high stretchability to enable dexterous flexibility and high stiffness to protect the musculoskeletal system. These two functions cannot be independently tuned in conventional engineering materials with linear or hyper- elasticity. Ligaments achieve this balance through a highly nonlinear tensile response characterized by a J-shaped curve, featuring an extended “toe region” of low force up to intermediate strains followed by an inflection, called the “heel region" which marks the onset of nonlinear stiffening. Here, we present a framework for characterizing the defining features of J-curve behavior.  Based on these features, we define measures for protectiveness and mobility to quantitatively describe the effective stiffness and the level of nonlinearity, thereby elucidating how the J-curve enables decoupled fine-tuning of flexibility and damage protection. A simplified mathematical model, supported by experimental validation, reveals the performance advantages of J-curve materials in antagonistic arrangements and highlights their unique design space compared with linear elastic systems. Furthermore, we develop synthetic J-curve materials capable of self-strain sensing via piezoresistive transduction, enabling their integration into practical devices. Collectively, these materials, models, and insights advance the understanding of nonlinear mechanical mechanisms in natural systems and provide a foundation for harnessing J-curve behavior in engineering applications such as bio-inspired robots.
\end{abstract}


\section{Introduction}

The biomechanics of musculoskeletal systems rely on the unique microstructural arrangements of both active and passive biomaterials from which complex mechanical characteristics emerge \cite{Meyers2013StructuralConnections}. Over 100 years ago, it was recognized that human muscles are nonlinear (see Figure 5 in \cite{HILL1953TheMuscle}). Similarly, ligaments and tendons exhibit J-shaped stress-strain curves \cite{Kwansa2010NovelRegeneration, L1978BIOMECHANICSTENDONS}. The nonlinear elastic behavior of muscles \cite{Evans1914TheMuscle}, tendons \cite{Maganaris1999InProperties}, and ligaments \cite{Fung1967ElasticityElongation, Woo1991TensileComplex} presents bioinspiration for novel engineering materials and devices \cite{Ma2017DesignReview}. The nonlinear elastic nature of ligaments results in a soft coupling between bones, thereby  \cite{Zhao2021BoostConstructs} providing a vital role in stabilizing joints, transmitting forces, and facilitating mobility. The biomechanical benefit of this J-shaped elastic curve is to maximize joint mobility while restraining excessive movements to protect against damage \cite{Blankevoort1988TheMotion}. Typically, we can use the maximum tensile force and the stretching elastic energy to refer to the protectiveness and the mobility, respectively. If we consider linear elastic materials, the mobility is quadratically proportional to the protectiveness. Specifically, if a linearly elastic ligament with spring constant of $k$ provides a resistive force $F_{max}$ within the natural range of motion of a joint, $d_{max}$, then it provides an internal resistive energy of $\frac{1}{2k}F_{max}^2$. However, such a proportional coupling constraint does not exist for J-curve materials. The nonlinear J-curve of natural ligaments enables independent tuning of mobility and protectiveness, not necessarily following the quadratic proportionality.

\begin{figure*}[t!]
\centering
\includegraphics[width=\textwidth]{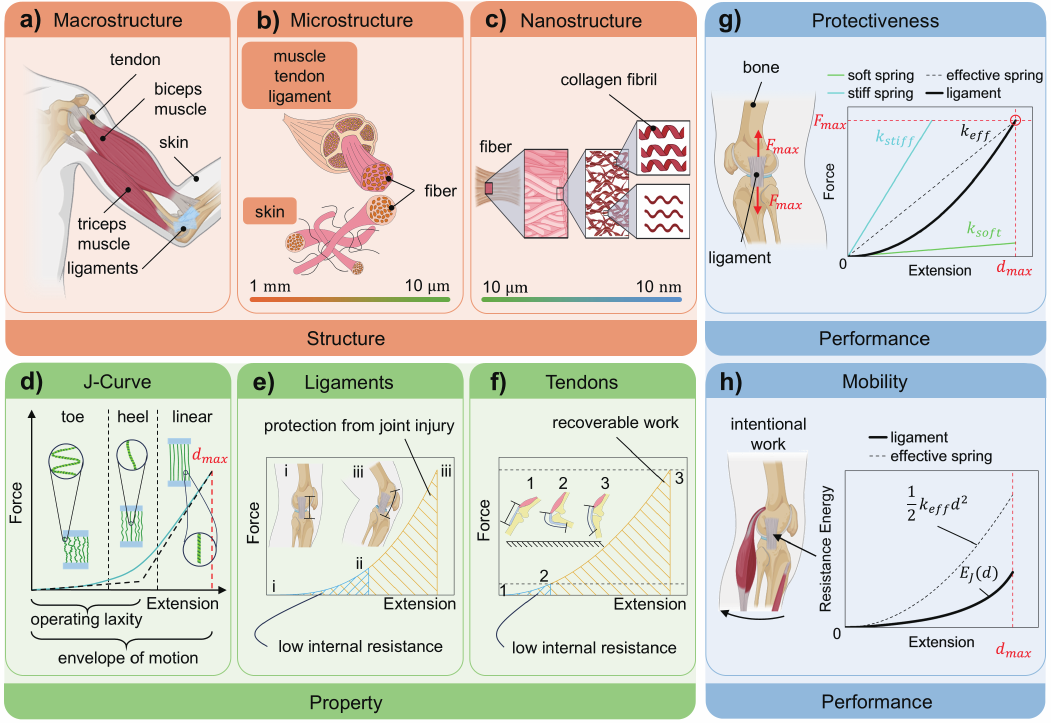}
\caption{Structure and properties of biological J-curve materials. a) Macroscale musculoskeletal structure of the human arm; b) The microstructures of muscle, tendon, ligament, and skin; c) The nanostructure of fibers in biological tissues; d) The representative tensile J-curve of biological tissues and illustrations for its mechanism; e) The property of ligaments with joint bending schematic; f) The property of tendons with an illustration of their role in the leg joint of a running turkey \cite{Roberts1997MuscularWork}; g) The protectiveness of the natural ligament under damaging force; and h) The mobility of the natural ligament quantified via the resistance energy.}
\label{fig:Lig_Fig1}
\end{figure*}

The complex hierarchical architecture of biological tissues is known to contribute to the J-curve \cite{Sasaki1996ElongationHierarchy, Kastelic1978TheTendon, EVANS1975StructuralFunction}. Natural muscles, tendons, ligaments, and skin have distinct shapes and functions, but all contain hierarchical fiber structures (Fig. \ref{fig:Lig_Fig1}a-b) with collagen fibrils serving as the fundamental element (Fig. \ref{fig:Lig_Fig1}c). Those nanoscale collagen fibers can be wavy (2D crimp) or twisted (3D helices) \cite{Safa2019HelicalTransfer, Sherman2015TheCollagen, Provenzano2006CollagenTendon, Lee2017ApplicationStudies}, and give rise to the J-curve \cite{Comninou1976DependenceFibres}. The tensile J-curve typically exhibits four phases. In the first phase of the force-displacement response (“toe region"), the collagen fibrils unfold so that they span a large strain region with bending-dominated deformation and ultra-low effective modulus (usually lower than a few tens of kPa for natural ligaments \cite{Huang2023BiomechanicalReconstruction}). With continued stretching, a “heel region" marks the onset of a second phase, where twisting and rotation of fibrils align the fibrils to the direction of applied stress and cause a distinct increase in modulus. In the third phase (“linear region"), the stretching of fibrils dominates the linear tensile response, resulting in an effective modulus several orders of magnitude larger than that of the first phase. The last phase is the “fracture region", where the tensile response reaches the maximum elastic stress, yield, and rupture.

Synthetic J-curve materials are inspired by this recent understanding of hierarchical architecture. Those synthetic materials with tensile J-curves are classified into 5 categories \cite{Ma2017DesignReview, Hanif2024RecentConstructs}: (i) planar network designs \cite{Jang2015SoftDesigns}; (ii) planar wavy and wrinkled designs \cite{Khang2006ASubstrates, Sun2006ControlledElectronics, Choi2007BiaxiallyNanomembranes}; (iii) three-dimensional (3D) helical designs \cite{Gerbode2012HowOverwinds, Chen2015HierarchicallyVapours, Abhari2018UsingYarns}; (iv) planar folding via kirigami \cite{Shyu2015ADefects, Isobe2016InitialMaterials} and origami \cite{Song2014OrigamiBatteries}; (v) 3D woven and knitted textile designs \cite{Maziz2017KnittingMuscles, Stoppa2014WearableReview}. Clearly, the common design motif among these materials is the straightening of waves, bends, or curls, leading to a low tensile modulus at a low strain level and a subsequent J-curve under additional strain. Several studies have investigated the effect of geometry on the J-curve, but these measurements were not related to the performance of the materials in practical use, nor to their function in bio-inspired antagonistic arrangements.

In this study, we investigate the features and performances of J-curve materials following the structure-property-performance framework. The objective is to study the characteristic features of the tensile behavior of J-curve materials and the unique advantages of these features over linear elastic materials in bio-inspired antagonistic arrangements. To help establish the theory, we develop novel synthetic ligaments by twisting and coiling elastomeric fibers with multiple structural levels, such as a primary structure (e.g., a helix) and a secondary structure (e.g., a ply), to form hierarchical architectures. We establish a systematic characterization framework to quantify the defining features of J-curves and demonstrate its universality by benchmarking against existing synthetic J-curve materials. Furthermore, we investigate the behavior of J-curve materials in antagonistic configurations and introduce a mathematical model that captures how pre-stretch transforms nonlinearity into stiffness, enabling the prediction of force output under various pre-stretch conditions. The model is validated through experimental testing of a linear antagonistic assembly constructed from two J-curve elements. To bridge toward practical applications, we also integrate electrical conductivity into the synthetic J-curve materials to enable self-strain sensing. Collectively, these results deepen our understanding of tensile J-curve phenomena by elucidating the fundamental bio-informed microstructural mechanisms that distinguish linear elastic and J-curve materials, and establish the principles of use of J-curve architectures in robotic applications.

\begin{figure*}[t!]
\centering
\includegraphics[width=\textwidth]{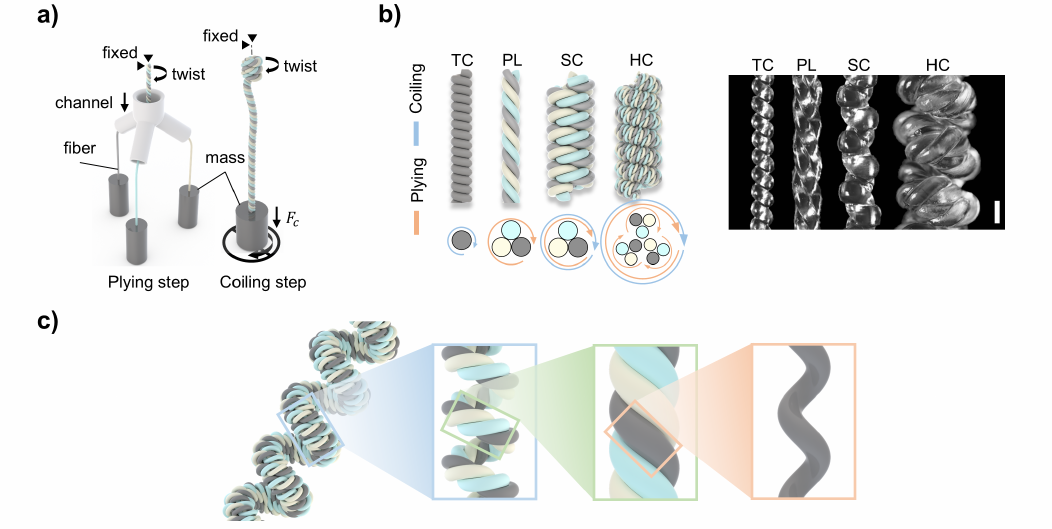}
\caption{Synthetic J-curve materials inspired by natural ligaments. a) Fabrication steps of making twisted and coiled artificial ligaments (TCALs); b) Computer-aided design (CAD) models and optical images of different types of TCALs fabricated with polyurethane (PU) fibers (scale bar is 1 mm). TC: Twisted and coiled, PL: Plied, SC: Supercoiled, and HC: Hypercoiled; and c) The hierarchical architecture ranking of artificial ligaments.}
\label{fig:Lig_Fig2}
\end{figure*}

\section{Properties and Performance of J-Curve Materials}

As shown in Fig. \ref{fig:Lig_Fig1}a-c, biological tissues in the musculoskeletal system have hierarchical architectures formed by numerous collagen fibrils with wavy and twisted arrangements. These unique structures contribute to the J-curve tensile behavior, as illustrated in Fig. \ref{fig:Lig_Fig1}d. For the entire envelope of motion of the J-curve, the value of the tangent slope monotonically increases with extension due to the stiffening effect of fiber microstructure transitioning from slack to straightened. The J-curve can be roughly divided into two parts with a transition point at the intersection of the tangent lines from the toe and the linear regions. This method is used to measure the laxity range of muscles, tendons, and ligaments in biomechanics. The properties of the two parts are distinct from each other and vary in the overall shape of the J-curves. For instance, differences in the J-curves of biological ligaments and tendons are attributable to variations in composition and structure \cite{Galbusera2022LigamentBiomechanics}. In Fig. \ref{fig:Lig_Fig1}e, the initial J-curve toe region has a wide extension range, providing the necessary flexibility to allow the bones in a joint to move freely within their natural range of motion. The resistive force in the linear region of the J-curve increases dramatically to limit joint motion and provide protection from injury. The features of the two J-curve regions result in natural ligaments that are both flexible and protective. Often, ligaments should not resist small motions within the joint. In contrast, the heel and linear regions of tendons play a critical role in storing considerable elastic energy, which improves movement efficiency by reducing muscle work and amplifying power \cite{Biewener2008TendonsFunction}. An interesting study \cite{Roberts1997MuscularWork} examined the states of muscle and tendon at the rear of the turkey's leg during running and found that the tendon's J-curve profile plays a crucial role in enhancing the turkey's running ability. Figure \ref{fig:Lig_Fig1}f contains a schematic of a turkey's leg during running, where running can be simplified as swing and stance phases. The elastic energy stored in the rear tendon is more than 60 \% of the muscle-tendon unit work, thus demonstrating the energy-recovery property of natural tendons. Besides, numerous examples have demonstrated the efficiency of movement in both human \cite{Lai2014TendonSpeed, Albracht2013Exercise-inducedHumans} and animal \cite{Biewener1998Muscle-tendonHorse} locomotion, wherein tendon extension provides recoverable work during gait.

\begin{figure*}[t!]
\centering
\includegraphics[width=\textwidth]{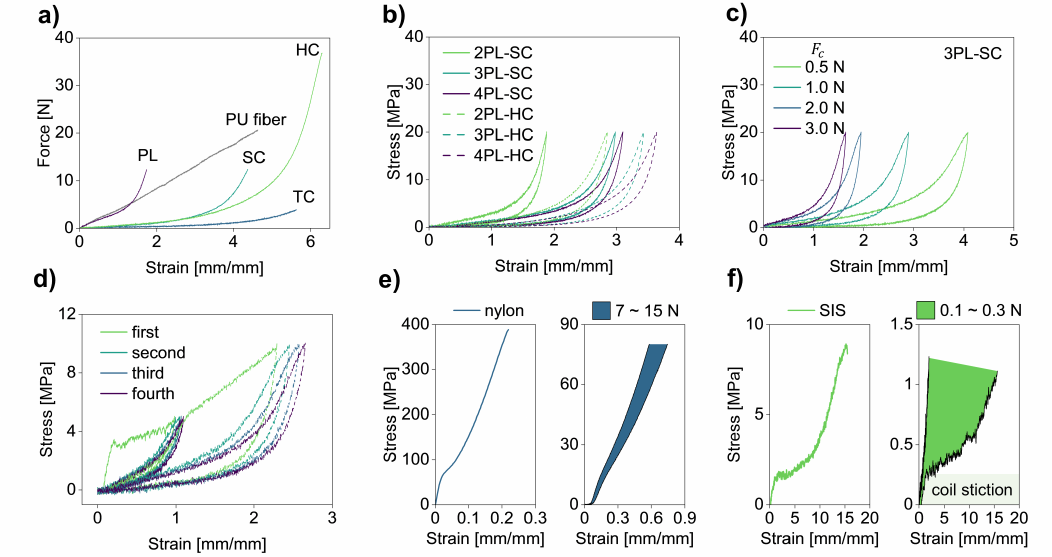}
\caption{Mechanical behavior of artificial ligaments. a) Tensile behavior of PU precursor fibers and PU fabricated TCALs; b-c) Experimental tensile loading and unloading of PU-fabricated TCALs. The former is focused on hierarchical TCAL variants, while the latter is only evaluating the coiling force effect on three-plied supercoiled (3PL-SC) artificial ligaments; d) Hysteresis and stress-dependent features of TCAL during four training cyclic tests at two incremental stress levels; and e-f) Tensile behavior of nylon and polystyrene-block-polyisoprene-block-polystyrene (SIS) fibers, as well as 3PL-SC artificial ligaments made by those two materials. Extreme coiling forces are applied to demonstrate tunability.}
\label{fig:Lig_Fig3}
\end{figure*}

The performance metrics of J-curve materials considered in this study are protectiveness and mobility, evaluated in two representative J-curve materials, ligaments and tendons. Protectiveness performance measures the J-curve effective stiffness (N/m) across the entire envelope of motion.  This linearization is used directly to compare J-curve materials with linear elastic materials, such as springs. An example is given in Fig. \ref{fig:Lig_Fig1}g, wherein a ligament connecting two rigid bodies is deformed under a force $F_{max}$. The force $F_{max}$ represents the critical force of the linear region within the envelope of motion. The line from the origin to the critical force is the tensile curve of the equivalent effective spring. The slope of this line, $k_{eff}$, is the stiffness of the J-curve and describes the deformation resistance ability. A spring with high stiffness $k_{stiff}$ reaches $F_{max}$ within a very small extension, which inhibits mobility or flexibility of joints, whereas a spring with low stiffness $k_{soft}$ provides greater flexibility but might never reach $F_{max}$ within the allowable range of motion. The second J-curve performance metric, mobility, measures the nonlinearity of the J-curve and is presented in the form of resistance energy. In Fig. \ref{fig:Lig_Fig1}h, the joint bends under intentional work while the antagonistic ligament provides the resistance energy, which is resisting the intended joint bending movement with a cost of energy. Within the allowable envelope of motion, we compare the resistance energy of the J-curve material to that of its effective spring. Theoretically, the resistance energy of a J-curve material follows the relationship equation \ref{equation: JcurveMobilityRange} along the envelope of motion:

\begin{equation} \label{equation: JcurveMobilityRange}
{\ 0 \leq E_J(d) \leq \frac{1}{2}k_{eff}d^2, \quad d \in [0, d_{max}]}
\end{equation}

where $E_J(d)$ is the resistance energy of J-curve function and $d_{max}$ is the critical displacement of the envelope of motion.

The resistance energy of the J-curve is always smaller than that of its effective spring over the entire envelope of motion due to the concave upward shape. We use the ratio of the resistance energy of the effective spring to that of the J-curve material as a measure of mobility performance. This measurement captures the geometric properties of the J-curve and provides an intuitive measure of the function of nonlinearity.

\section{Bio-inspired Synthetic J-Curve Materials}

We constructed synthetic J-curve materials to study their mechanical function in antagonistic arrangement, inspired by the architecture of biological J-curve materials shown in Fig. \ref{fig:Lig_Fig1}a–c. In this section, we introduce a family of synthetic J-curve materials made from twisted and coiled polymer fibers with bio-inspired structures. To evaluate the reliability of this bio-inspired approach, we examined the tensile behavior of the synthetic J-curve materials, the effects of manufacturing parameters, the hysteresis phenomenon, and the feasibility of multi-materials.

\subsection{Fabrication and Structure of Artificial Ligaments}

A series of twisted and coiled artificial ligaments (TCAL) with complex, unique geometries was fabricated by applying twists and coils on polymer materials such as polyurethane (PU) fibers through plying \& coiling. Fig. \ref{fig:Lig_Fig2}a shows a schematic that describes the strategies applied to TCALs. The plying step replicates the twisted fiber arrangements in biological J-curve materials by twisting three parallel fibers together. Guiding channels are used in the plying step to guarantee order during the plying process. The coiling step aims to create a uniform helical structure where each fiber is directly connected to a non-rotatable weight (providing coiling force $F_c$) at the bottom. The fabrication method involves different combinations of these two steps, each producing a unique type of TCAL. Mainly, four types of TCALs (twisted \& coiled (TC), plied (PL), supercoiled (SC), and hypercoiled (HC)) are fabricated using those two fabrication steps on PU fibers. The schematic and optical images in Fig. \ref{fig:Lig_Fig2}b illustrate the structure of all TCALs, along with diagrams that highlight the combinations of fabrication steps for each TCAL. All four types of TCALs draw inspiration from the structural characteristics of biological ligaments. The TC and PL designs directly mimic the helical and twisted fiber arrangements found in these tissues, and the SC structure combines both. Whereas the HC structure extends the concept by incorporating hierarchical architecture atop combinations of wavy and twisted structures. Fig. \ref{fig:Lig_Fig2}c is the hierarchical structure level ranking of TCALs to better illustrate the relationship between different TCAL types.

\subsection{Mechanical Behavior and Tunability}

To investigate the relationship between TCAL geometries and their mechanical behaviors, tensile tests were conducted on all types of TCALs using a Universal Testing Machine (Instron) with details described in the \textit{SI Appendix}. The tensile responses of PU precursor fiber and various kinds of PU-made TCALs are plotted in Fig. \ref{fig:Lig_Fig3}a to visualize their discrepancies in tensile behaviors. This figure plots force versus engineering strain, eliminating the effect of initial length, and suggests that a nonlinear tensile response exists across all types of TCALs, demonstrating the feasibility of using the plying \& coiling method to replicate the nonlinear tensile performance of natural ligaments. Additionally, we observed that the modulus of PL in the initial portion is significantly higher than that of TC, SC, and HC, which is related to differences in their primary structures. The screw axis of the latter three TCALs is helical and differs from the straight axis of PL, leading to discrepancies in their toe region modulus. At the same strain level, wavy structures like a helix are naturally more slack than twist structures and hence require less force to be stretched. We also compared the TCAL modulus differences at the ends and found that the modulus of the end portion is related to the number of fibers used to fabricate TCALs. For TC, it has only one PU fiber, so its modulus at the end portion is similar to that of the single PU fiber. For PL and SC, they are all made of three PU fibers and hence have a modulus that is approximately three times that of the PU fiber. As for HC, it contains nine PU fibers and has the largest end portion modulus among all tensile curves. However, due to the effect of complex fiber interactions, hysteresis, and creep, their modulus in the force versus strain plot is not linearly proportional to the number of fibers. Those results demonstrate the feasibility of using TCAL to replicate the mechanical behavior of natural ligaments and show that TCAL's mechanical properties are influenced by factors such as structures and fiber number.

The mechanical behavior tunability of TCALs results from their structure, number of fibers, and fabrication process parameters. Fig. \ref{fig:Lig_Fig3}a presents the tuning effect of different structures and suggests the potential influence of the number of fibers. For TCAL, varying the number of fibers alters the geometric distribution of stress among fibers, thereby altering the tensile response. To further investigate the influence of the number of fibers, we prepared variants of SC and HC structures with 2 or 4 fibers at their twisted structures (the detailed schematic and optical images are shown in \textit{SI Appendix}, Fig. S1). The experimental results for all variants are shown in Fig. \ref{fig:Lig_Fig3}b, including both tensile loading and unloading curves to show the full cycle. The figure is plotted as a function of engineering stress (defined as the tensile load divided by the cross-sectional area of all fibers) and strain, with the final stress limited to 20 MPa in all cases. Hence, by maintaining the same maximum stress, we compare the end strains of each sample. We found that TCAL is more stretchable with more fibers. Also, increasing the hierarchical architecture levels increases TCAL stretchability, as evidenced by the end strains of SC and HC with the same number of fibers. Besides, the increase in TCAL stretchability becomes smaller as the number of fibers increases. Those results show the effect of the fiber number on the stretchability of the TCAL tensile response. 

The mechanical behavior can also be altered by modifying the fabrication process parameters such as the twist density, defined as the total rotational radians divided by the original length of the precursor fiber. The twist density represents the torsional strain applied to fibers. According to the Kirchhoff-Love rod theory \cite{F.1920AElasticity}, such twists require a stretching force to sustain, which is the tension applied during fabrication. With the help of tension, the twist inserted is maintained, forcing the polymer chain to elongate along the twisted direction and making the fiber stiffer \cite{Wang2024TheActuators}. This implies that TCAL's stretchability is qualitatively related to tension and can consequently be controlled by varying the fabrication mass. To validate the hypothesis, we fabricated 3-ply supercoiled (3PL-SC) artificial ligament samples with four different coiling forces ($F_c$) and tested their tensile loading and unloading behavior at a constant end stress of 20 MPa. In Fig. \ref{fig:Lig_Fig3}c, the stretchability of samples decreases with the coiling force, which is consistent with the prediction of the hypothesis. Notably, the end strain of J-curves in Fig. \ref{fig:Lig_Fig3}c spans a wide range (1.5-4.2), indicating that the coiling force has a significant effect on the tensile behavior of TCAL. Additionally, the decrease in end strain becomes smaller with increasing coiling force. These results demonstrate the feasibility of developing a practical strategy to tune the nonlinear tensile response of TCALs to meet specific performance requirements. 

The complex fiber-fiber interactions in TCALs exhibit hysteresis and stress-dependent behavior, which we control via a preconditioning process. To illustrate the effect of the preconditioning process, we conducted two continuous cyclic tests at different stress levels, each repeated 4 times. As Fig. \ref{fig:Lig_Fig3}d shows, the first tensile test of each cycle is clearly different from the following 3 cycles. This phenomenon is primarily caused by internal deformation of the TCAL structure during the first tensile cycle, which is associated with the rearrangement of fiber orientations. Such fiber orientation rearrangement, which mainly occurs during the first tensile test, is responsible for the obtained J-curve behavior in the subsequent cycles. The difference between the tensile response decays from cycle 2 to cycle 4 and eventually leads to a steady tensile J-curve behavior. Additionally, the fiber orientations of TCALs are influenced by the tensile test stress level, which we control by specifying a particular end stress for samples made from a certain material. In Fig. \ref{fig:Lig_Fig3}d, the tensile response of TCALs at higher stress levels differs substantially from that observed at lower stress levels. These observations highlight the critical role of the tensile testing protocol and preconditioning procedure in assessing the influence of tuning parameters, as variations in testing methods can lead to inconsistent outcomes even for identical samples. To ensure stability and reproducibility, we established a standard tensile testing procedure that incorporates preconditioning and training steps to mitigate hysteresis. This protocol was applied uniformly to all TCAL samples to maintain consistency and avoid stress-dependent artifacts. A detailed description of the method is provided in the \textit{SI Appendix}.

Multiple materials were fabricated into TCALs and their tensile behaviors were tested, including polyamide 6 (nylon, a thermoplastic polymer used in twisted \& coiled polymer actuators) and polystyrene-block-polyisoprene-block-polystyrene (SIS, an ultra-soft viscoelastic polymer). To utilize those materials as fibers for TCAL manufacturing, we developed a spinning method to produce custom SIS fibers. The details of the spinning method are described in the materials and methods section and \textit{SI Appendix}. The tensile curve of each material is shown on the left side of Fig. \ref{fig:Lig_Fig3}e and Fig. \ref{fig:Lig_Fig3}f, and each fiber was stretched until fracture. From those plots, the variations between the materials are very distinct (Young's modulus for each material: nylon is 2.3 GPa, PU is 26 MPa, and SIS is 0.63 MPa), which is sufficient to validate the feasibility of creating J-curves with different materials. To compare with the tensile results of PU in Fig. \ref{fig:Lig_Fig3}c, nylon and SIS fibers were fabricated into the 3PL-SC structure with varying coiling force to explore their tunability. In addition, the end-stress levels for TCAL samples were adjusted based on each material property. For the various materials tested, the ratio between the selected end stress and the fiber ultimate stress ranged from 10 \% to 20 \%. The tensile test results for all TCAL samples are given on the right side of Fig. \ref{fig:Lig_Fig3}e and Fig. \ref{fig:Lig_Fig3}f. From the plotted results, all TCAL tensile curves exhibit nonlinear behavior and differ substantially from the tensile response of the base materials within the same stress range, demonstrating the feasibility of replicating J-curve behavior across different materials using our fabrication method. The nylon TCAL samples display a typical J-curve at low strains. However, we note that at 0.2 strain we see non-monotonic changes in their tangent slopes, which could be indicative of the effect of hysteresis. Moreover, nylon TCALs have a smaller strain window than PU, suggesting that the PU TCAL benefits from complex interaction between the hyperelasticity of the fibers, their interactions, and the coiled architecture. In addition, SIS-based TCAL samples also exhibit a steep initial slope, attributable to coil stiction arising from the inherently tacky surface of SIS fibers. These observations indicate that although the fabrication method can induce J-curve–like behavior across diverse materials, the intrinsic properties of each material remain the dominant factor governing the tensile response.

\begin{figure*}[t!]
\centering
\includegraphics[width=\textwidth]{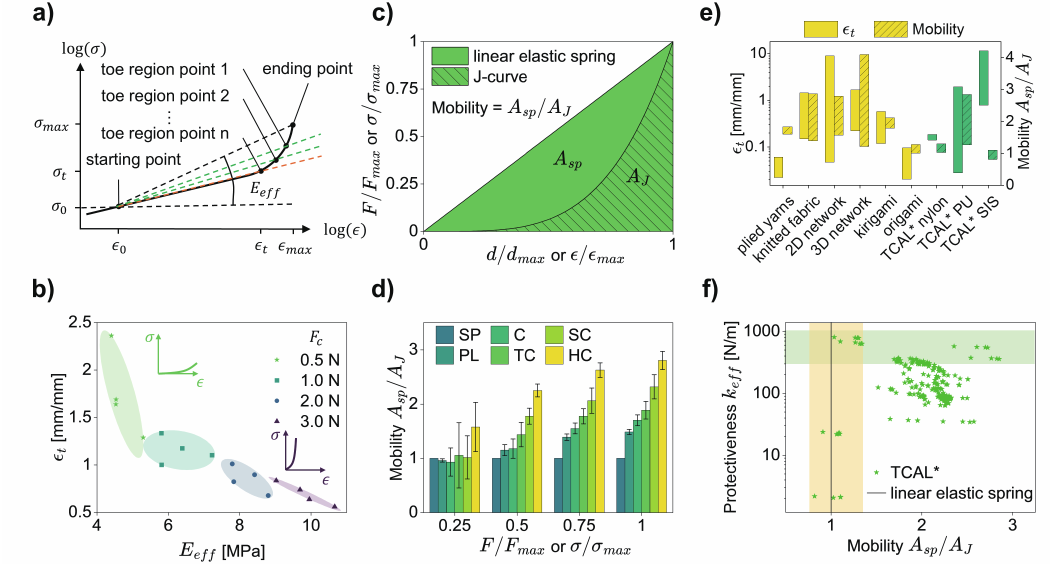}
\caption{Features of J-curve. a) An iteration method that determines the toe region zone and the effective modulus of the whole J-curve; b) J-curve shape evaluation method with the toe-region strain and effective modulus using the coiling force effect on TCAL as an example; c) The definition of mobility to evaluate the nonlinearity of the J-curve; d) The mobility corresponding to several stress levels with experimental results from the linear elastic spring and PU-made TCALs; e) The toe-region strain and mobility of existing synthetic J-curve materials and multi-material TCALs. Non-hatched bars are showing the toe-region strain, while the hatched bars are presenting the mobility; and f) The Ashby chart compares linear elastic materials and J-curve materials using experimental data from all TCAL samples. The yellow region illustrates the range of protectiveness that can be tuned in J-curve materials while maintaining constant mobility, whereas the green region highlights the mobility tunability achievable at a fixed level of protectiveness.}
\label{fig:Lig_Fig4}
\end{figure*}

\section{Principles of Use of J-Curve}

Next, we define and measure the features that capture the characteristics of J-curve materials and their performance in practical scenarios. 

The entire section is divided into three parts. The first part characterizes the J-curve feature, including definitions and comparisons across different J-curve materials. The second part introduces the principles for using J-curve materials in antagonistic arrangements and benchmarks their advantages over linear elastic materials. The final part integrates a strain-sensing function into the synthetic ligaments, making them more practical for real engineering applications.

\subsection{Feature Characterization of J-Curve}

The characteristic of the J-curve defines the displacement and forces within the toe, the heel, and the linear regions. In biomechanics, the toe region is often described as the initial stage of tendon or ligament deformation in which the crimp pattern of collagen fibers is straightened \cite{Wang2006MechanobiologyTendon}, typically occurring within the first 2 \% strain. However, this definition is not universally applicable, as many J-curve materials do not exhibit a crimp structure, and hence the 2 \% strain only applies to the specific J-curve. Here, we develop a reproducible definition of the toe region based on an iterative fitting method. First, we use logarithmic scales with base 10 on both axes to enhance the visibility of the toe region behavior, as shown in Fig. \ref{fig:Lig_Fig4}a. Then, we filter the noise below a minimum threshold stress $\sigma_0$ at the low-strain region. In the second step, the entire loading curve, from the starting point to the ending point, is used to fit a line that begins at the starting point. Since this is a log-log plot, the line is a power law, with the slope being defined by the exponent. The cross-point of the fitting line and the loading curve is named the toe region point 1. Point 1 is always at a smaller strain than the original endpoint, and hence defines a new, narrower strain range.  A new fit is then generated using only the data from the starting point to the first toe region point. This iterative process continues, updating the toe region point in each cycle, until the difference between successive toe region points falls below a specified threshold:

\begin{equation} 
{\delta_k \equiv |\vec{v_{k-1}}-\vec{v_{k}}| \leq \alpha, \quad k = 2,..., n}
\label{equation: toeregioniteration}
\end{equation}

Here, the $\delta_k$ represents the difference between two vector points, $\alpha$ is the threshold used in this method (the specific value used is), and $\vec{v_k}$ is a vector that contains the stress and strain values of the toe region point k. Once this inequality is fulfilled, the iteration ends, and the last point is the toe region point in this definition. This approach thus defines a power-law fit for the toe region, with a fitted exponent ranging from 0.55 to 2.05 for TCALs. 

Based on this definition, numerical metrics can be used to assess and benchmark J-curves. Those parameters include $\epsilon_t$ and $\sigma_t$ (strain and stress values of the toe region point) and the effective modulus $E_{eff}$ (the linear slope of the stress-strain curve calculated from the starting point to the ending point at normal scale). Those parameters characterize the shape of the J-curve and can thus be used to evaluate the effects of the process parameters on tuning the J-curve. Fig. \ref{fig:Lig_Fig4}b illustrates the impact of coiling forces—previously identified as changes in stretchability—on PU-made 3PL-SC synthetic ligaments by $\epsilon_t$ and $E_{eff}$. The plot shows that as the coiling force increases, $\epsilon_t$ decreases while $E_{eff}$ increases, with $\epsilon_t$ and $E_{eff}$ representing the stretchability and stiffness of the J-curve, respectively. These trends indicate that tuning the coiling force inherently imposes a trade-off between stretchability and stiffness in the tensile J-curve of TCALs.

We define non-dimensional mobility as a measure of the J-curve's nonlinearity. The new definition is shown in Fig. \ref{fig:Lig_Fig4}c, where the mobility is defined as the area ratio of the effective linear elastic spring and the J-curve. In this case, mobility is defined as a dimensionless term emphasizing the difference between the J-curve and the straight slope line of the equivalent spring. In Fig. \ref{fig:Lig_Fig1}h, the mobility was described in light of the elastic resistance energy of the J-curve, but an energy value cannot clearly measure the level of the nonlinearity of the J-curve. On the other hand, the new definition facilitates direct comparison of J-curves across different axis conventions (e.g., stress–strain, force–displacement, or mixed plots). By normalizing each curve to its respective maximum value, as shown in Fig. \ref{fig:Lig_Fig4}c, the shapes and degrees of nonlinearity of different J-curves can be compared more clearly and consistently.

To demonstrate the use of mobility to compare various TCALs, we compare the mobility of a spring (SP) with that of various TCAL structures at different stress levels by normalizing the J-curves below the stress levels. Those TCAL structures include PL, coiled (C), TC, SC, and HC, where the difference between C and TC is that C has only pure geometric coiling, without any internal twists in the fiber. Besides, those different types of TCAL structures are fabricated in a manner (details are included in \textit{SI Appendix}, Fig. S3 and Table S1) so that they have the same spring index, allowing them to be comparable in mechanical behavior. In Fig. \ref{fig:Lig_Fig4}d, the HC structure has the highest mobility among all structures, showing that a higher level of hierarchy provides advantages to the J-curve nonlinearity. Additionally, the mobility increases with the stress levels due to the proximity to the heel region of the J-curve. 

\begin{figure*}[t!]
\centering
\includegraphics[width=\textwidth]{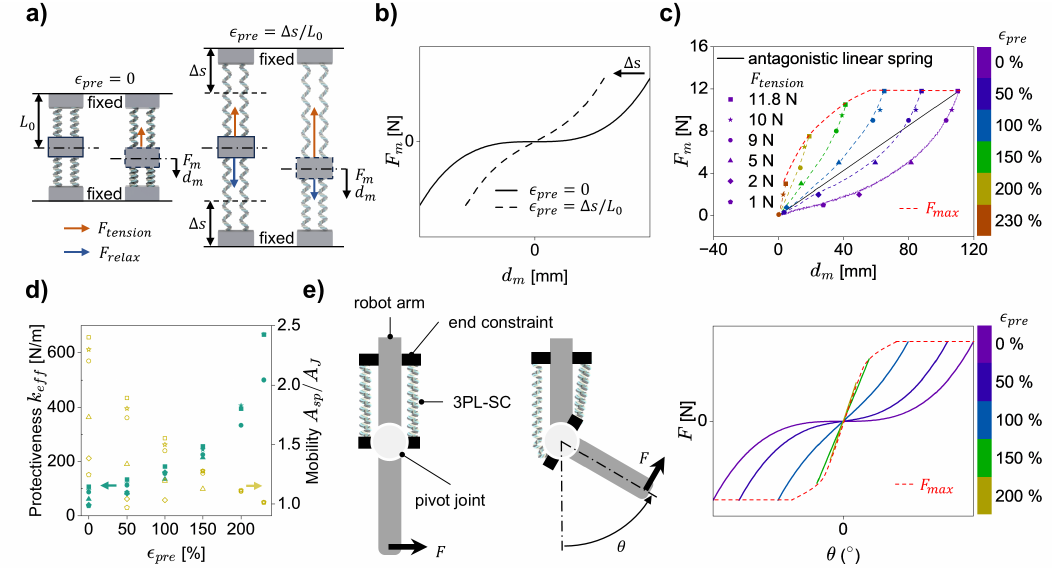}
\caption{Principles of use of J-curve materials in antagonistic arrangements. a) Theoretical force-displacement predictions of linear antagonistic structure with \& without pre-stretch; b) Schematics of the linear antagonistic structure with \& without pre-stretch; c) Results of linear antagonistic test with theoretical and experimental data. Theoretical predictions are plotted in dashed curves (the zero pre-stretch curve is from tensile experiments), and experimental data are plotted in dots. A line of equivalent antagonistic linear spring systems is also given for comparison. The color bar shows the pre-stretch level, and different dot shapes indicate the total tension force $F_{tension}$ of TCALs. The red dashed curve represents the damaging force at different pre-stretch levels; d) The exchange of the protectiveness and mobility values for the linear antagonistic structure with different $F_{tension}$ and pre-stretch strains. Orange solid dots are the change of protectiveness, while purple hollow dots are the alteration of mobility; and e) Schematics of the rotary antagonistic structure with a pivot joint. A normal external force is applied at the end of the lower arm to rotate the lower arm with angle $\theta$. The right side shows the predicted result of the rotary antagonistic structure.}
\label{fig:Lig_Fig5}
\end{figure*}

To verify that the proposed J-curve characterization methods are applicable across different types of J-curve materials, we evaluated existing synthetic J-curve systems using the toe-region strain and mobility metrics to compare their stretchability and nonlinearity. The work we evaluated contains most of the synthetic J-curve material types, including i) plied yarns \cite{PinedaGuzman2021ReplicationYarn}, ii) knitted fabric \cite{Jang2014RuggedMonitoring, Kononova2011Modelling3950}, iii) 2D network \cite{Jang2015SoftDesigns, Lanzara2010AMaterials, Chen2015StatusLigaments}, iv) 3D network \cite{Jiang2016Highly-stretchableMetamaterials, Yan2020SoftDesigns}, v) kirigami \cite{Bahamon2016GrapheneArrays, Jiang2016Highly-stretchableMetamaterials, Qi2014AtomisticKirigami}, and vi) origami \cite{Silverberg2015OrigamiFreedom, Dudte2016ProgrammingOrigamitessellations}. The benchmark is shown in Fig. \ref{fig:Lig_Fig4}e, where we compare the experimental results of existing synthetic J-curve materials with those of the TCAL made with PU, nylon, and SIS. This benchmark validates the universality of the feature-extraction method shown in Fig. \ref{fig:Lig_Fig4}a and c. Furthermore, this benchmark provides an intuitive evaluation of the tunability of each study by visualizing the range of toe-region strain and mobility. From the plot, we can see that different types of synthetic J-curve materials exhibit distinct mobility ranges. We inferred that such mobility depends on the deformation geometry of each structure. For example, the 3D network has the largest mobility range among all types of J-curve materials. The allowable degree of freedom, and hence the design space, is higher for 3D compared to 2D network types, which contributes to the differences in their mobility ranges. Hence, with this benchmark, we show that the TCAL covers nearly the entire range of previously reported toe-region strain and mobility. From a practical perspective, the manufacturing method for TCAL is scalable, involving only twisting and coiling of polymer fibers, and their fiber-like form factor makes them suitable for integration into robotic joints inspired by biological ligaments and tendons. 

Within the structure-property-performance framework, we focus on the protectiveness and mobility, as defined above. We plot an Ashby chart to compare the performance differences between linear elastic materials and J-curve materials in terms of protectiveness and mobility. In Fig. \ref{fig:Lig_Fig4}f, the experimental data from all TCAL samples are compared against the performance of linear elastic springs. Because linear elastic springs exhibit a strictly linear tensile response within their elastic regime, their mobility is fixed at 1, resulting in a vertical line on the plot. In contrast, J-curve materials occupy a broader region of the design space due to their nonlinear tensile behavior, which affords additional degrees of freedom. This tunability enables J-curve systems not only to vary protectiveness while maintaining constant mobility—analogous to a linear spring—but also to modulate mobility while holding protectiveness constant, as illustrated by the yellow and green regions in Fig. \ref{fig:Lig_Fig4}f.

\subsection{Antagonistic Arrangements for J-Curve Materials}

The presence of the antagonistic arrangement is quite common in nature. The upper arm, with the biceps and triceps muscles arranged antagonistically, is a classic example from human anatomy. Such an arrangement allows the elbow to be flexed and extended with the assistance of the biceps and triceps. The primary disadvantage of the antagonistic arrangement is the unavoidable passive stretching of one side when the other is flexed, resulting in greater resistance than in a single-muscle arrangement. Nonlinear tensile performance is required on both sides to reduce resistance at moderate flexing. In such a case, the resistance for flexing the elbow is vastly decreased, mainly with the help of the J-curve, and only poses considerable resistance under large-angle flexing. This example shows the advantage of J-curve materials against linear elastic materials in increasing the flexibility of the antagonistic structure. 

We can simplify this antagonistic arrangement into a linear antagonistic arrangement model, depicted in Fig. \ref{fig:Lig_Fig5}a, where two identical TCAL groups are connected in series with fixed ends. External force $F_m$ is exerted on the middle block in the axial direction, leading to the corresponding displacements $d_m$ of the middle block. The force-displacement profile $F_m(d_m)$ of the middle block is the output of the entire antagonistic system. The tension forces from the top and bottom TCAL groups are identified as $F_{tension}$ and $F_{relax}$, respectively, based on the direction of $F_m$. For zero pre-stretch level ($\epsilon_{pre} = 0$), all TCAL groups are at their original length $L_0$. The theoretical prediction of the system output at zero pre-stretch level is shown in Fig. \ref{fig:Lig_Fig5}b, which is essentially the sum of the mechanical behaviors from the top and the bottom TCAL groups. To simplify the model, we reduce the full mechanical behavior to piecewise functions $\mathcal{F}(d)$ and limit the range within the critical displacement of the envelope of motion $d_{max}$:

\begin{equation} \label{equation: Superposition_zeroPrestretch}
    \begin{gathered}
    F_m(d_m) = \mathcal{F}(d_m) - \mathcal{F}(-d_m)
    \\ 
    \mathcal{F}(d) = 
    \begin{cases}
        \mathcal{J}(d), & 0 \leq d \leq d_{max}\\ 
        0, & d < 0 
    \end{cases}
    \end{gathered}
\end{equation}

Here, the two TCAL groups are identical, so that their mechanical behaviors $\mathcal{F}(d)$ are the same with opposite deformation $d$ and $-d$. The $\mathcal{J}(d)$ represents the J-curve function of the TCAL group at the tensile test. Based on the equation \ref{equation: Superposition_zeroPrestretch} and the Fig. \ref{fig:Lig_Fig5}b, the system output at each direction is equal to the tensile mechanical behavior of the single TCAL group at zero pre-stretch level. We experimentally validated this prediction using the linear antagonistic setup, as described in the materials and methods section. The results are shown in Fig. \ref{fig:Lig_Fig5}c, where the purple solid curve is the tensile experimental result of a single TCAL group and all the purple dots are the experimental results from the linear antagonistic setup. Different dot geometries stand for different $F_{tension}$. From the plot, we can see that the test results align well with our prediction at the zero pre-stretch level, and compared with the linear elastic spring system (black solid line), the former has significant flexibility. 

Another advantage of J-curves in antagonistic arrangements is the tunable stiffness of the system by pre-stretching J-curve materials, which contributes to the resilience of the musculoskeletal system to dynamic loads. While challenging to measure, the existence of pre-stretch in ligaments is known to be critical to providing joint stability and allows for nearly instantaneous response to external perturbations \cite{Shirazi-Adl2006EffectForces, Kersh2015TheLigaments}.  Without pre-stretch, joint stability would not be enforced until system slack is removed. For example, ensuring the appropriate amount of pre-stretch is critical for successful ligament reconstruction surgeries and re-establishing proper joint kinematics \cite{Nicholas2004AReconstruction}. The mechanisms of ligament and tendon pre-stretch span are complex and change during locomotion.  The macroscopic orientation of the tissue provides a resistive line of action that changes during joint actuation, resulting in non-uniform loading of the tendon or ligament. The intrinsic microstructure (e.g., crimp angle, degree of twist, and density) of both fibrils and fibers has developed to create a pre-tensile state suited and tuned for each joint's function. 

For linear elastic springs, the system output is a straight line with twice the stiffness of the single spring group. The system stiffness cannot be changed by pre-stretching each spring group, and details are discussed in the \textit{SI Appendix}. 

For J-curve systems, the prediction of system stiffness with pre-stretch strain ($\epsilon_{pre} = \Delta s/L_0$) is given in Fig. \ref{fig:Lig_Fig5}b, which is the summation of shifted mechanical behavior of the top and bottom TCAL groups. Consequently, we re-express the system output $F_m(d_m)$ of J-curve materials as:

\begin{equation} \label{equation: Superposition_Prestretch}
    \begin{gathered}
    F_m(d_m) = \mathcal{F}(d_m + \Delta s) - \mathcal{F}(-d_m + \Delta s)\\
    0 \leq \Delta s \leq d_{max}
    \\
    \end{gathered}
\end{equation}

where $\Delta s$ is the initial stretch applied to each TCAL group. Because of the nonlinear tensile function $\mathcal{J}(d)$, the system output becomes stiffer than the zero pre-stretch case, with less nonlinearity and stretchability. By applying different pre-stretches to this equation, we compute prediction curves for the different pre-stretch strain levels, shown as the dashed curves (except for the red dashed curve) in Fig. \ref{fig:Lig_Fig5}c. For TCAL groups, the maximum $F_{tension}$ they can have is defined as the damaging force $F_{max}$, and it is presented as the end force of each curve $F_m(d_{max})$. Note that the $F_{max}$ (red dashed curve) starts decreasing with the pre-stretch level when the $d_{max}$ of a certain pre-stretch level is less than half of the $d_{max}$ at zero pre-stretch. This is because the external force $F_m(d_m)$ is equal to the sum of $F_{tension}$ and $F_{relax}$ for all pre-stretch levels:

\begin{equation} \label{equation: Antagonistic}
    \begin{gathered}
    F_m(d_m) = F_{tension} + F_{relax}
    \\
    \end{gathered}
\end{equation}

Note that $F_{tension}$ and $F_{relax}$ have opposite directions to each other. Therefore, at high pre-stretch levels (more than half of the zero pre-stretch $d_m$), the $F_{tension}$ would be larger than the external force $F_m$ as part of the force is provided by the $F_{relax}$ from the other side. Since we are not specifying the value of $F_{tension}$, this equation applies to all conditions with different $F_{tension}$. To validate this, we conducted experiments on several $F_{tension}$ levels, with each level labeled with a different data marker in Fig. \ref{fig:Lig_Fig5}c. From the plot, the dots from the same pre-stretch level match the prediction curve, and the dots from the same $F_{tension}$ level all follow the same decreasing trend as the $F_{max}$. The plateau portion of the $F_{max}$ curve indicates that the equivalent modulus of the whole system can increase within this portion without causing a decrease in $F_{max}$. Interestingly, engineering measures such as safety factors can be introduced to expand its mechanical applications and enhance safety by choosing a lower $F_{tension}$ since the plateau portion applies to all $F_{tension}$. 

With increased pre-stretch levels, the J-curve system becomes stiffer and more closely resembles linear springs. From Fig. \ref{fig:Lig_Fig5}d, we can see that there exists a trade-off between protectiveness and mobility with pre-stretch strain levels, indicating that the pre-stretch fine-tunes the mobility and the protectiveness in such a way that when the mobility increases, the protectiveness value decreases. This exchange between the two performance metrics, merely due to assembly in antagonistic arrangements, offers J-curve systems an extraordinary wide range of applications in engineering, as they can not only provide high mobility, like biological ligaments in joints, but also be tuned to behave as linear elastic springs and supply significant stiffness.

In addition to the linear antagonistic arrangement, we also conceptually evaluated the rotary antagonistic arrangement, which is more common in biological and engineered systems. The left side of Fig. \ref{fig:Lig_Fig5}e shows a schematic of the rotary antagonistic structure with a pivot joint and robot arms. The TCAL groups are placed on either side of the robot arm and connect to the pivot joint to provide passive resistance. External force $F$ is applied to the normal direction of the lower arm to rotate the pivot joint with angle $\theta$, stretching the left side TCAL group. The right side of Fig. \ref{fig:Lig_Fig5}e illustrates the system output predictions of the rotary antagonistic structure, where the x variable is changed to the rotation angle $\theta$. In this case, the displacement $d_m$ is the arc length change due to rotation, which is proportional to the rotation angle $\theta$. The fundamental mechanism of this rotary antagonistic arrangement is the same as the linear antagonistic arrangement, thus applicable to the same principles.

\begin{figure*}[t!]
\centering
\includegraphics[width=\textwidth]{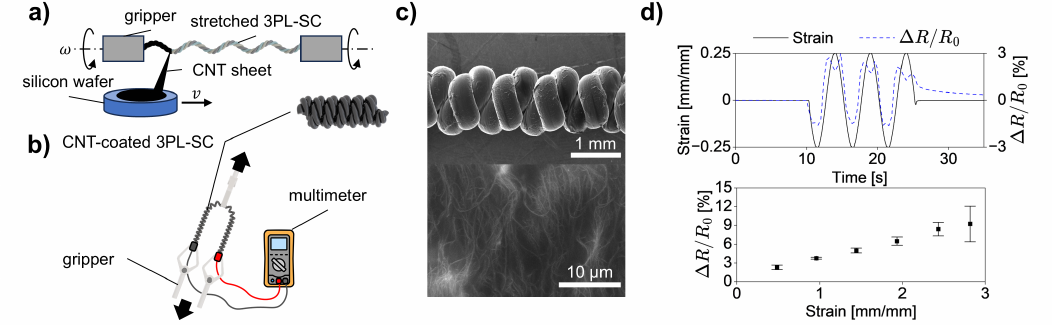}
\caption{Self-sensing J-curve TCAL. a) The coating method of carbon nanotube (CNT) coated 3PL-SC artificial ligament; b) The strain-sensing test setup of CNT-coated 3PL-SC artificial ligament; c) Scanning electron microscope (SEM) images of the CNT-coated 3PL-SC artificial ligament; and d) Strain-sensing results of CNT-coated 3PL-SC artificial ligament. The top figure shows the relative change in resistance during the cyclic tensile test, with the black solid line representing the strain change and the blue dashed line showing the relative change in resistance. The bottom figure shows the peak-to-trough relative change in resistance with different fluctuating strain wave amplitudes.}
\label{fig:Lig_Fig6}
\end{figure*}

\subsection{Self-sensing Application}

The adjustable stretchability and excellent nonlinearity make the J-curve material, such as TCAL, suitable for mechanical applications. However, it also means that using an independent strain sensor to measure TCAL strain is challenging, especially for measuring large deformations. Therefore, TCAL self-sensing would provide valuable information for strain control when TCALs are used in robotics. To make TCAL self-sensing, we use the electrically conductive coating with piezoresistive transduction capability. There are several existing piezoresistive coatings \cite{Tsai2022ElectricallyDeposition, Kuwabara2017SubstratePolymerization, Akter2012ReversiblyNanowires} and approaches for mixing the polymer with conductive liquids \cite{Haque2020ProgrammableComposites} or conductive fibers \cite{Yang2023ElectrospunPerspectives}. For this study, we coated a thin layer of carbon nanotubes (CNTs) on the surface of the 3PL-SC artificial ligaments, making them conductive while preserving their stretchability. For the coating method, 3PL-SC samples were stretched (approximately two times their initial length) and mounted on grippers of a unique setup that can rotate both grippers in the same direction with the same speed $\omega$ (Fig. \ref{fig:Lig_Fig6}a). The CNT sheet was attached to the stretched 3PL-SC sample surface. The rotation of grippers and the linear motion (with speed $v$) of the silicon wafer worked simultaneously to coat the CNT sheet on the surface of the 3PL-SC sample. Detailed coating procedures are provided in the materials and methods section. SEM images of the prepared 3PL-SC sample are shown in Fig. \ref{fig:Lig_Fig6}c.

After coating CNT on 3PL-SC samples, the strain-sensing performance was evaluated by conducting a cyclic tensile test while monitoring the electrical resistance. The Fig. \ref{fig:Lig_Fig6}b shows the test setup. In this figure, the CNT-coated 3PL-SC sample is bent in a “U" shape, with both terminals attached with probes connected to a multimeter to monitor the resistance change. The “U" shape helps eliminate the resistance change due to the motion of the probes. In this case, both probes are placed on the bottom side. Grippers connected to both ends remain stationary, and the gripper connected to the curved region moves with the tensile stage to perform the tensile test without affecting the contact resistance of the probes. The results of the test are given in Fig. \ref{fig:Lig_Fig6}d, where the top plot shows the strain changes, as well as the corresponding relative change of resistance. The bottom plot illustrates the relationship between the relative change of resistance and the amplitude of the strain wave. From those plots, it is clear that the resistance of the CNT-coated 3PL-SC artificial ligament increases with strain, indicating its ability to detect strain changes up to a strain of 3. With its large range of strain-detection capabilities, high stretchability, and protective function, the CNT-coated TCALs can serve as a powerful tool across various applications.

\section{Discussion}

This work defined quantitative characteristic features to benchmark the tensile J-curve of bio-inspired synthetic ligaments and tendons.  Within the structure-property-performance framework, we investigated the principles of using J-curve materials and their performance when used in antagonistic arrangements. The study used these measured features and the performance benchmarks to demonstrate that J-curve materials have potential for broader engineering applications than linear elastic materials. By analyzing the tensile J-curve feature and the properties of biological ligaments and tendons in the joint motion, we concluded that the performance of the J-curve are protectiveness and mobility. Protectiveness is the overall stiffness of the J-curve and can thus be directly compared with that of linear elastic springs. As for mobility, it is represented as the non-dimensional measure of energy that resists the intentional motion and is smaller than the effective linear spring with the same stiffness. In addition, mobility provides an intuitive way to evaluate the nonlinearity of the J-curve.

Inspired by the twist-and-wavy structure of natural ligaments, we developed a plying \& coiling processing method to fabricate polymer fibers into synthetic materials that can replicate the tensile J-curve property. We validated the reliability of this bio-inspired approach by duplicating tensile J-curves across the four main variants of the synthetic J-curve material and demonstrated how material structure and manufacturing parameters affect stretchability tuning. Due to the hysteresis and stress-dependent features, the tensile behavior of synthetic J-curve materials is highly dependent on the tensile test method. To this end, we developed a systematic tensile test method to eliminate the influence of the test method and maintain the consistency and reliability of the results. The feasibility of this method is also validated across multiple materials, but their material properties largely constrain the J-curve feature.

The features of the J-curve include a series of parameters which describe the geometric attributes of the J-curve, including the protectiveness and critical point of the toe region. Those parameters reflect the stiffness and stretchability of the J-curve, and can thus be used to indicate the effects of tuning factors. The latter is defined as the mobility, a non-dimensional parameter that captures the comparable nonlinear feature across all J-curves. This quantitatively defines the essential difference between the linear springs and the higher degrees of design freedom present in the J-curve materials. Certain features of the J-curve are excluded from this work. Bilinear materials with wrinkled structures \cite{Ma2016DesignElectronics} exhibit performance comparable to that of J-curve materials, but suffer from sharp transitions, which could cause serious dynamic issues when used as tendons in robotic applications due to instantaneous high acceleration near the transition point. Therefore, the smoothness feature might be added to the evaluation of J-curve materials in the future.

We discovered that the antagonistic structures with J-curve units exchange the protectiveness and the mobility values using pre-stretch. As demonstrated through a linear antagonistic structure model and validation experiments, the pre-stretch applied to the J-curve units stiffens the entire system, which reduces the intrinsic range of nonlinearity of the system compared to the units. The force output of the antagonistic system is the summation of tension from the antagonistic pair of J-curve units. Therefore, for a specific J-curve unit tension, the entire system can be stiffened up to twice its effective stiffness without decreasing the maximum force output. In addition, these principles can be applied to the rotary joint antagonistic structure, thereby significantly expanding its range of applications. We recognize that the dynamic behavior of J-curve systems, including their response time and frequency-dependent performance, is crucial for real-world applications and should be the subject of future work. 

This study establishes a framework for understanding, analyzing, and leveraging biological and engineered materials that exhibit tensile J-curve behavior. By examining the performance of J-curve materials in antagonistic configurations, we reveal fundamental advantages over traditional linear systems and underscore the broad potential of J-curve architectures in engineered applications. Looking ahead, we anticipate that an increasing number of bio-inspired robots and engineered systems will incorporate J-curve materials to enhance dexterity, adaptability, and overall functional performance, especially for bio-inspired and humanoid robots.  Moreover, the investigation of novel synthetic J-curve materials provides new insights into biological J-curve materials, especially those in antagonistic arrangements in the musculoskeletal systems.

\section{Methods}

\subsection{Materials}

For nylon, the 0.47 mm diameter monofilament of PowerMono Fishing Line is purchased from RUNCL. For PU, several monofilaments are purchased from Beadalon, including those with diameters of 0.5 mm, 0.8 mm, and 1 mm. The particles of SIS are purchased from Sigma-Aldrich with product number 432393-500G.

\subsection{SIS Fiber Spinning Method}

Dissolve styrene-isoprene-styrene (SIS) particles with toluene (50 wt\%) in a single-neck round-bottom flask with a thermal bath (oil bath for 6 hours at 120 $^{\circ}\text{C}$) and keep stirring the solution (30 rpm). The round-bottom flask was connected to a Graham condenser to prevent toluene evaporation and maintain the SIS concentration. The solution was then cooled to room temperature. Pour the SIS solution into the syringe and seal both ends. Remove air bubbles by using a planetary mixer (AR-100, Thinky) and place the nozzle with a 19-gauge needle on the syringe's front end. Fill a tank with acetone and submerge the nozzle in it. Constant air pressure (10 psi) was used to push the SIS solution out of the nozzle and submerge it into acetone at a slow and steady pace. Take the SIS fiber out when it changes from transparent to white. Dry the white SIS fiber in the oven (40 $^{\circ}\text{C}$, 5 mins) to get rid of acetone. The color should turn transparent when it is thoroughly dried. Detailed explanation of the mechanism is included in the \textit{SI Appendix}.

\subsection{Artificial Ligament Fabrication}

The fabrication methods of artificial ligaments follow a similar pattern, and thus, choosing the supercoiled artificial ligament as an example. In the plying step, cut the required number of fibers (three for 3PL-SC) into segments of the same length and clamp with a ring terminal (7113K91, McMaster-Carr). Secure the ring terminal to the rotational motor and have each fiber pass through the corresponding channels. Ply the clamped fibers with constant rotational speed from the top motor while keeping fibers in tension by applying weight at the free end. Tether the unclamped end of all three fibers with a ring terminal after plying. For the coiling step, hang the weight at the bottom end to provide tension and constrain the rotation of the bottom end. Twist the plied fibers again with a constant rotational speed to make the plied fibers bend and form coils. Stop twisting when it is fully coiled. Mount the supercoiled ligament on a frame to hold its geometry and then anneal it.

\subsection{Annealing Method}

The TCAL manufactured from fabrication methods must be thermally annealed to set its geometry. The annealing setup uses a hot plate to heat peanut oil or distilled water, allowing TCALs to be annealed uniformly. Nylon is annealed in an oil bath at 170 $^{\circ}\text{C}$ for 2 hours. PU uses the same oil bath annealing and time at 120 $^{\circ}\text{C}$. SIS uses a water bath with 1 \% weight of tetrahydrofuran (THF) annealing at 50 $^{\circ}\text{C}$ for 1 hour.

\subsection{Linear Antagonistic Test}

TCAL samples for the antagonistic test were pre-treated with a cured epoxy block (High Heat Epoxy Syringe Dark Grey, J-B Weld) at both terminals. Two parallel TCALs were fixed together to form a single unit, and two units were used to create the antagonistic structure. Both TCAL units were mounted on an 80/20 rail with their terminals secured and connected at the middle block (as shown in \textit{SI Appendix}, Fig. S7). The middle block is attached with a cable for load application. For the top TCAL unit, the top end is connected to a force sensor (LSB200, FUTEK Advanced Sensor Technology Inc.). This test has pre-stretch level from 0 \%, 50 \%, 100 \%, 150 \%, 200 \%, and 230 \% for both TCAL units. Payloads were applied to the middle block at each pre-stretch level to maintain a top unit tension force of 1 N, 2 N, 5 N, 9 N, 10 N, and 11.8 N, respectively. The force sensor can read the tension force of the top unit, and the corresponding displacement of the middle block was recorded. The tensile force caused solely by pre-stretch is called the initial tensile force, and any tensile force setting lower than this initial value is discarded. Besides, the lower TCAL unit will be removed if its length is shorter than its initial length to eliminate any compression-related influence.

\subsection{Carbon nanotube Coating Method}

A carbon nanotube (CNT) dry coating tool was developed to wrap the TCAL with CNT sheets drawn directly from CNT forests synthesized on the silicon substrate (\textit{SI Appendix}, Fig. S8). This tool features two grippers that can rotate simultaneously, allowing the TCAL attached to them to rotate accordingly. To use this tool, the TCAL is cut to the proper length and fixed by both grippers. Then, tweezers are used to pick up a CNT strip from the silicon wafer and place it on the TCAL. Rotating the tool handle allows TCAL to rotate and have the CNT strip fully coated on TCAL. The CNT wafer is gradually translated in the axial direction of TCAL while rotating the TCAL. This process consistently coats the surface of TCAL with CNT until it reaches the other end. To add more layers, we change the translation direction of the CNT substrate and continue rotating the handle until the CNT strip returns to its initial position. This wraps the TCAL with two layers of CNT. Finally, we cut the extra CNT strip and dismounted the CNT-coated TCAL from the tool. We soak the CNT-coated TCAL with isopropyl alcohol to condense the CNT on the surface of the TCAL using capillary forces.

\section*{Data Availability}
All data is included in the article and/or \textit{SI Appendix}

\section*{Acknowledgments}
We acknowledge funding from the United States Office of Naval Research N00014–22–1–2569, and Grainger College of Engineering Strategic Research Initiative. Portions of the figures were created with BioRender.

\bibliographystyle{unsrtnat}
\bibliography{references}

\end{document}